\documentclass[letterpaper]{article} 
\pdfoutput=1
\usepackage[submission]{aaai23}  
\usepackage{times}  
\usepackage{helvet}  
\usepackage{courier}  
\usepackage[hyphens]{url}  
\usepackage{graphicx} 
\usepackage{caption} 
\usepackage{cite}
\usepackage{subcaption}
\DeclareGraphicsExtensions{.eps}
\usepackage{algorithm}
\usepackage{algorithmic}
\usepackage[dvips]{epsfig}  
\usepackage{amssymb}
\usepackage{amsmath}
\usepackage{newfloat}
\usepackage{listings}
\usepackage{bibentry}
\DeclareCaptionStyle{ruled}{labelfont=normalfont,labelsep=colon,strut=off} 
\floatstyle{ruled}
\newfloat{listing}{tb}{lst}{}
\floatname{listing}{Listing}
\title{A Generative Adversarial Network Approach for Identification and Mitigation of Cyber-Attacks in Wide-Area Control of Power Systems}

\author{
     by AAAI Press Staff\textsuperscript{\rm 1}\thanks{With help from the AAAI Publications Committee.}\\
    AAAI Style Contributions by Pater Patel Schneider,
    Sunil Issar,\\
    J. Scott Penberthy,
    George Ferguson,
    Hans Guesgen,
    Francisco Cruz\equalcontrib,
    Marc Pujol-Gonzalez\equalcontrib
}

\title{Artificial Intelligence based Approach for Identification and Mitigation of Cyber-Attacks in Wide-Area Control of Power Systems}
\author {
    Jishnudeep Kar,\textsuperscript{\rm 1,2}
    Aranya Chakrabortty \textsuperscript{\rm 1,3}
}
\affiliations {
    \textsuperscript{\rm 1} North Carolina State University\\
   \textsuperscript{\rm 2}jkar@ncsu.edu, \textsuperscript{\rm 3}achakra2@ncsu.edu
}

\begin{document}

\maketitle
\begin{abstract}
We propose a generative adversarial network (GAN) based  deep learning method that serves the dual role of both identification and mitigation of cyber-attacks in wide-area damping control loops of power systems. Two specific types of attacks considered are false data injection and denial-of-service (DoS). Unlike existing methods, which are either model-based or model-free and yet require two separate learning modules for detection and mitigation leading to longer response times before clearing an attack, our deep learner incorporate both goals within the same integrated framework. A Long Short-Term Memory (LSTM) encoder-decoder based GAN is proposed that captures the temporal dynamics of the power system significantly more accurately than fully-connected GANs, thereby providing better accuracy and faster response for both goals. The method is validated using the IEEE 68-bus power system model. 
\end{abstract}

\section{Introduction}

\begin{figure*}[h]
    \centering
    \includegraphics[width=\linewidth,height=4.5cm]{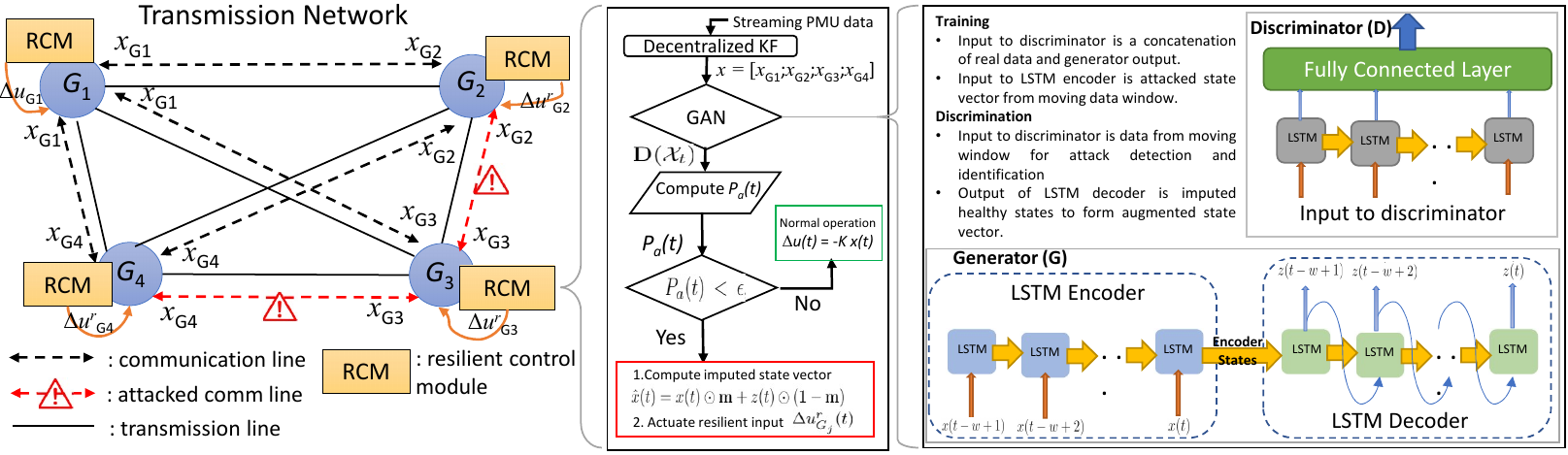}\vspace{-0.2cm}
    \caption{Proposed LSTM-GAN framework and associated learning architecture}
    \label{fig:gan}
\end{figure*}
Cyber-security solutions for wide-area control of power systems have garnered significant attention over the past decade \cite{cs2}. Two most common and pertinent types of attacks that have been addressed in the literature include false data injection (FDI) attacks, where intruders can hack in to the wide-area communication network channels and corrupt the data streams originating from Phasor Measurement Units (PMUs) \cite{fdi}, and Denial-of-Service (DoS) attacks where attackers may temporarily deactivate a link to  obstruct the PMU data to reach their desired destination \cite{dos}. Majority of the algorithms for detecting and mitigating these types of attacks reported so far are model-based, meaning that they require the knowledge of the exact power system model at the time of the attack. However, given the recent expansion in the number of PMUs, grid operators are gradually inclining towards more model-free and data-driven approaches \cite{liu} such as neural network-based deep learning solutions, as reflected in the recent works \cite{nn2}. In \cite{jkarnn}, we proposed a neural network based approach to identify a closed-loop control from controller bank to mitigate denial of service attacks. In \cite{jkarlstm}, a LSTM based approach was used to compensate for missing information due to a DoS attack. However, this was severely influenced by the operating point, and large changes in operating point caused the LSTM to create biased estimates. One major limitation of these approaches, however, is that the errors due to insufficient sampling in the training phase can cause inaccurate identification of the operating point of the grid, which, in turn, can lead to inaccurate control actions.

In this paper we propose a generative adversarial network (GAN) \cite{gan} based learning strategy which serves dual purposes of both identification and mitigation of DoS attacks and FDI attacks.  In our proposed framework, each GAN can be implemented in a decentralized manner at every generating station, thereby avoiding the need for any communication between the GANs. We use an encoder-decoder structure for the ``generator" module of the GAN using LSTM to accurately capture the grid dynamics. The input to the generator is the attacked state vector (containing either missing state entries for DoS, or corrupt state entries for FDI), and the output is the  GAN's best prediction for what these missing or anomalous entries should be in normal conditions. We  modify the loss function for training the GAN to ensure that the output of the generator closely resembles the input state vector for entries that are healthy. The predicted state vector is then fed to the ``discriminator" module of the GAN, which serves to discriminate between the real PMU data streaming in to the substations and the generated data. The novelty of our approach lies in the fact that we train a single integrated GAN model for both detection and mitigation that guarantees more accuracy, and also that the GANs can be installed independently at critical generating stations although they work collectively to protect the grid from incoming attacks. 

\section{Problem Statement} 
Consider a power system with $n$ synchronous generators. Every generator is modeled by a standard $7^{th}$-order dynamic model  $x_{G_i} = [\delta_i\;\omega_i\;\psi_{1di}\;E_{qi}\;\psi_{2qi}\;E_{di}\;E_{fdi}]^T$. The physical meaning of these states can be found in \cite{kundur} and are skipped here for brevity. The nonlinear dynamic model is written in the form
\begin{equation}
    \dot{x} = F(x) + Bu, \, x(0) = x_0
    \label{eqn:nonlinear}
\end{equation}
where, $x = [x_{G_1}^T \, x_{G_2}^T \,\hdots\, x_{G_n}^T]^T \in \mathbb{R}^{7n}$, $F(x) \in R^{7n}$ is a vector of nonlinear functions modeling the swing and excitation system dynamics \cite{kundur}, $B \in \mathbb{R}^{7n \times n}$ is the input matrix and $u = [u_{G_1}\; u_{G_2}\; .... \; u_{G_n}]^T \in \mathbb{R}^n$ is the excitation voltage, which is used as a control input actuated through the AVR/PSS circuit. We assume that generator states can be estimated at each generator using
the decentralized Kalman filters (UKF) \cite{kf} facilitated by strategically placed PMUs that guarantee geometric observability of the grid. The small signal model of (\ref{eqn:nonlinear}) can be written as 
\begin{equation}
    \Delta \dot{x} = A \Delta x + B\Delta u, \, \Delta x(0) = \Delta x_0,
    \label{eqn:lti}
\end{equation}
    where $\Delta x = x-x_o$, $x_o$ being the load-flow solution for desired operating point, ($A,B$) are small-signal state and input matrices of appropriate dimensions, and $\Delta u$ is the small-signal control input. The objective is to damp system-wide power oscillations, which is formulated using a linear quadratic regulator (LQR) based optimal controller with weights $Q \geq 0,\; R > 0$. A state-feedback controller $\Delta u = -K \Delta x$ is designed where $K$ minimizes the LQR objective constrained to (\ref{eqn:lti}). 
Every generator $i$ is equipped with the $i^{th}$ row of $K$ denoted by $K_{i,j}, \, j=1,..,n$, so that it computes its control signal at any time $t$ as $ \Delta u_{G_i}(t) = - \Big( \sum_{G_j, j=[1,..,n]} K_{i,j} \big( x_{G_j}(t) - x_{o,G_j} \big)   \Big)$, 
where $x_{o,G_j}$ from (\ref{eqn:lti}) is the linearization point for generator $j, j=1,..,n$. Every generator needs to communicate with every other generator to share state information as shown in Fig. \ref{fig:gan}. We consider that a malicious attacker can attack one or more communication links using a FDI or DoS attack, leading to an incorrect control input  given by\vspace{-0.1cm} \scriptsize
$
      \Delta u^a_{G_i}(t) = - \Big( \sum_{G_j \in \mathcal{S}_i} K_{i,j} \big( x_{G_j}(t) - x_{o,G_j} \big) +  \sum_{G_k \in \mathcal{A}_i} K_{i,j} \big( x^a_{G_k}(t) - x_{o,G_k} \big)  \Big),\vspace{-0.05cm}
$\normalsize
where $\mathcal{S}_i$ is the set of generators from whom $G_i$ is receiving healthy states,  $\mathcal{A}_i$ is the set of generators from whom $G_i$ is receiving compromised or no states, and $x^a_{G_k}(t)$ represents the attacked vector.\vspace{-0.2cm}

\section{Proposed GAN Design Framework}
\subsection{1. Implementation architecture}
In (2), the $A$ matrix may change depending on the operating conditions such as loads and generation dispatches. Therefore, our proposed GAN method must work for a variety of possible operating conditions.  As shown in Fig. \ref{fig:gan}, we propose to implement a GAN at every generator (or, a subset of critical generators that have highest influence on the inter-area oscillations). It takes as input the instantaneous state estimate that it receives from the Kalman filters of other generators over the communication network, and identifies an attack, followed by generating values for the attacked entries which can be used for correctly actuating the control input. The purpose of the GAN is two-fold : 1) detect and identify an attack, 2) redesign the control $\Delta u(t)$ after the attack is cleared. The discriminator $\mathbf{D}$ is used to detect any FDI, whereas the generator $\mathbf{G}$ is used to predict the actual values of the attacked entries to construct an imputed state vector to actuate the modified  control input. Note, the data here is the streaming time series measurements of the estimated generator states by the UKF located at the generator buses, which do the estimation based on the PMU measurements (current, voltage phasors) from geometrically observable set of buses.

\subsection{2. Redesigned training of the GAN}
 The generator $\mathbf{G}$ consists of a LSTM based encoder-decoder structure. The encoder takes the incomplete state vector as input over a time window of size $w$. The LSTM states of the encoder are further fed to the decoder LSTM which then outputs the corresponding generated state vector. The discriminator $\mathbf{D}$ takes the output of the generator as its input, and outputs a probability score indicating if the input is real data or data from the generator. In general, the training of a GAN can be formulated as \cite{gan} \vspace{-0.1cm}
\begin{equation}
    \min_{\theta_{\mathbf{G}}} \max_{\theta_{\mathbf{D}}} L(\mathbf{D},\mathbf{G}) = \mathbb{E}[\log \mathbf{D}(\mathcal{X})] + \mathbb{E} [1 - \log \big( \mathbf{D} (\mathbf{G}(\mathcal{Z})) \big)],
\end{equation}\normalsize
where, $\mathcal{X}$ denotes the real-data, $\mathcal{Z}$ denotes the input to the generator,  $\mathbb{E}$ denotes the expectation, $\theta_{\mathbf{G}}$ and $\theta_{\mathbf{D}}$ are the generator and discriminator networks parameters, respectively. However, training the GAN using simply the above loss function will not be sufficient for our problem as for a given input of attacked states $\mathcal{Z}$, the generator would output any $\mathcal{X}$ that maximizes the discriminator score without considering the healthy entries in the input. Moreover, we also want the output of the generator to be close to the input for states belonging to $\mathcal{S}_i$. Therefore, we define a reconstruction loss $L_r$ as $L_r (\mathbf{G}) = || \big( \mathcal{Z} - \mathbf{G}(\mathcal{Z}) \big) \odot \Omega ||,
    \label{eqn:reconsloss}$, where $\Omega$ is a binary matrix of 0's for attacked entries and 1's for healthy states. Next, we re-define the training problem as \vspace{-0.1cm}
\begin{equation}
    \min_{\theta_{\mathbf{G}}} \max_{\theta_{\mathbf{D}}} L(\mathbf{D},\mathbf{G}) + \lambda L_r(\mathbf{G}),
    \label{eqn:modifiedloss}\vspace{-0.2cm}
\end{equation}
where $\lambda$ is used to calculate a weighted sum and can be chosen as a suitable hyper-parameter. The detailed flow-diagram for the training steps is in Figure 1. \vspace{-0.1cm}

\subsection{3. Usage for online identification and resiliency} 
\subsubsection{Detection and identification of cyber-attack : } 
Consider a snapshot of the states for a window size $w$. Denote the associated measurement matrix for $G_j, j=1,..,n$ at time $t$ as
$
    \mathcal{X}^{G_j}_t = \text{col}\big(x({t-w+1}) , x({t-w+2}) , . . . ,  x({t}) \big)
$,  where $\mbox{col}(.)$ represents a matrix with its embraced vectors as the columns, and with slight abuse of notation, $x(t)$ represents the states received by $G_j$. In the ideal case, for detection we  want to have $\mathbf{D}(\mathcal{X}^{G_j}_t)   < \epsilon , $  if  $ t>t_0 $, and $\mathbf{D}(\mathcal{X}^{G_j}_t)   > \epsilon , \, $ if $t< t_0$ ,
where $\epsilon>0$ (considered as 0.5 in simulations later) is a threshold, then the data could be deemed as compromised. However, since the trained GAN may not be perfect, there may be individual instances where this rule may be violated because of the prediction error. This would raise a false alarm and lead to unnecessary control actions for mitigation. Instead, we propose a moving average method, wherein the probability of attack $P_a(t)$ is given by
\begin{equation}
    P^{G_j}_a(t) = \big( \mathbf{D}(\mathcal{X}^{G_j}_t) + \mathbf{D}(\mathcal{X}^{G_j}_{t-1}) + \hdots + \mathbf{D}(\mathcal{X}^{G_j}_{t-d+1}) \big) / d,
    \label{eqn:Pa}\vspace{-0.1cm}
\end{equation}
 where, $d$ denotes the moving average window. 
At any time, if $P_a(t) < \epsilon$, then an attack is detected, and an alarm is raised. As we will see in results, the advantage of using the moving average is that it does not change significantly due to anomalous individual discriminator error.

For the identification step, we use the entry-wise reconstruction loss. Let us denote $L_r^{G_j}(t)$ as the reconstruction loss corresponding to the entry in $\mathcal{X}_t$ for $j^{th}$ generator as\vspace{-0.11cm}
\begin{equation}
   L_r^{G_j}(t) = || \big(\mathcal{X}_t - \mathbf{G}(\mathcal{X}_t)\big)  \odot \Omega_{G_j} ||.
   \label{eqn:LGi}\vspace{-0.11cm}
\end{equation}
If for any time $t$, the value of $L_r^{G_i}(t) > \epsilon_{G_i}$, where $\epsilon_{G_i}$ is a corresponding threshold value, then the data corresponding to $G_i$ is deemed as compromised. 
\begin{algorithm}[h]
  \begin{algorithmic}[1]
\STATE \textbf{For each} $t \in [0 \;,\; T]$
\STATE \,\, Compute input matrix $\mathcal{X}^{G_j}_t$
\STATE \,\, Calculate the discriminator score   $\mathbf{D}(\mathcal{X}^{G_j}_t)$ and $P^{G_j}_a(t)$
\STATE \,\, If $P^{G_j}_a(t) < \epsilon$, then attack is detected and 
\STATE \,\,\,\,\, \textbf{For each} $i=1, 2, \hdots, n$
\STATE \,\,\,\,\,\,\, Compute $L_r^{G_i}(t)$ according to (\ref{eqn:LGi})
\STATE \,\,\,\,\,\,\, If $L_r^{G_i} > \epsilon_{G_i}$
\STATE \,\,\,\,\,\,\,\,\, Add $G_i$ to the attack set $\mathcal{A}_j$ for generator $G_j$
\STATE \,\,\,\,\,\,\, Else
\STATE \,\,\,\,\,\,\,\,\, Add $G_i$ to secure set $\mathcal{S}_j$ for generator $G_j$
\STATE \,\,\,\,\,\,\, End if
\STATE  \textbf{End \{for, if, for\}}
  \end{algorithmic}
  \caption{Attack identification at $G_j, j=1,..,n$}
  \label{alg:identification}
\end{algorithm}

\subsubsection{Resiliency using GAN based imputation : }
Once the set of attacked generators $\mathcal{A}_j$ has been identified, we construct the attack binary vector $\mathbf{m}_{G_j}$. Let us represent $\mathbf{G}(\mathcal{X}^{G_{j_t}})$ as \vspace{-0.1cm}
\begin{equation}
    \mathbf{G}(\mathcal{X}^{G_j}_t) =\mbox{col}\big(z({t-w+1}) , z({t-w+2}) , . . . ,  z({t}) \big),
    \label{eqn:Z}\vspace{-0.05cm}
\end{equation}
where $z(t)$ represents the respective column vectors of  $\mathbf{G}(\mathcal{X}_t)$.
Next, to calculate the control from the combination of the healthy states and the generated states, we construct the imputed state vector as \vspace{-0.1cm}
\begin{equation}
    \hat{x}(t) = x(t) \odot \mathbf{m}_{G_j} + z(t) \odot (\mathbf{1} - \mathbf{m}_{G_j}).
    \label{eqn:imputation}\vspace{-0.1cm}
\end{equation}
where $\odot$ is the Hadamard product. Please note that we augment the healthy states with the predicted states only for the attacked entries. Therefore, the final control can be actuated as $\Delta u^r_{G_j}(t) = - K_j \hat{x}(t)$, where $K_i$ is the $i^{th}$ row of $K$.
\begin{algorithm}[h]
  \begin{algorithmic}[1]
\STATE \textbf{For each} $t \in [0 \; , \; T]$
\STATE \,\, Use Alg. \ref{alg:identification} to identify attack and generate sets $\mathcal{A}_j$, $\mathcal{S}_j$
\STATE \,\, \textbf{If attack detected}
\STATE \,\,\,\,  Using sets $\mathcal{A}_j$, $\mathcal{S}_j$, construct the binary vector $\Omega_{G_j}$
\STATE \,\,\,\, Compute the imputed state vector $\hat{x}(t)$ 
\STATE  \,\,\,\, Actuate the resilient control \footnotesize$\Delta u^r_{G_j}(t) = - K_j \hat{x}(t)$\normalsize
\STATE \,\, \textbf{Else} Go to step 2
\STATE  \textbf{End \{if , for\}}
  \end{algorithmic}
  \caption{Online implementation at $G_j, j=1,..,n$}
  \label{alg:online}
\end{algorithm}\vspace{-0.2cm}

\section{Numerical Results}
\begin{figure}[h]
    \centering
    \includegraphics[width=0.97\linewidth,height=4cm]{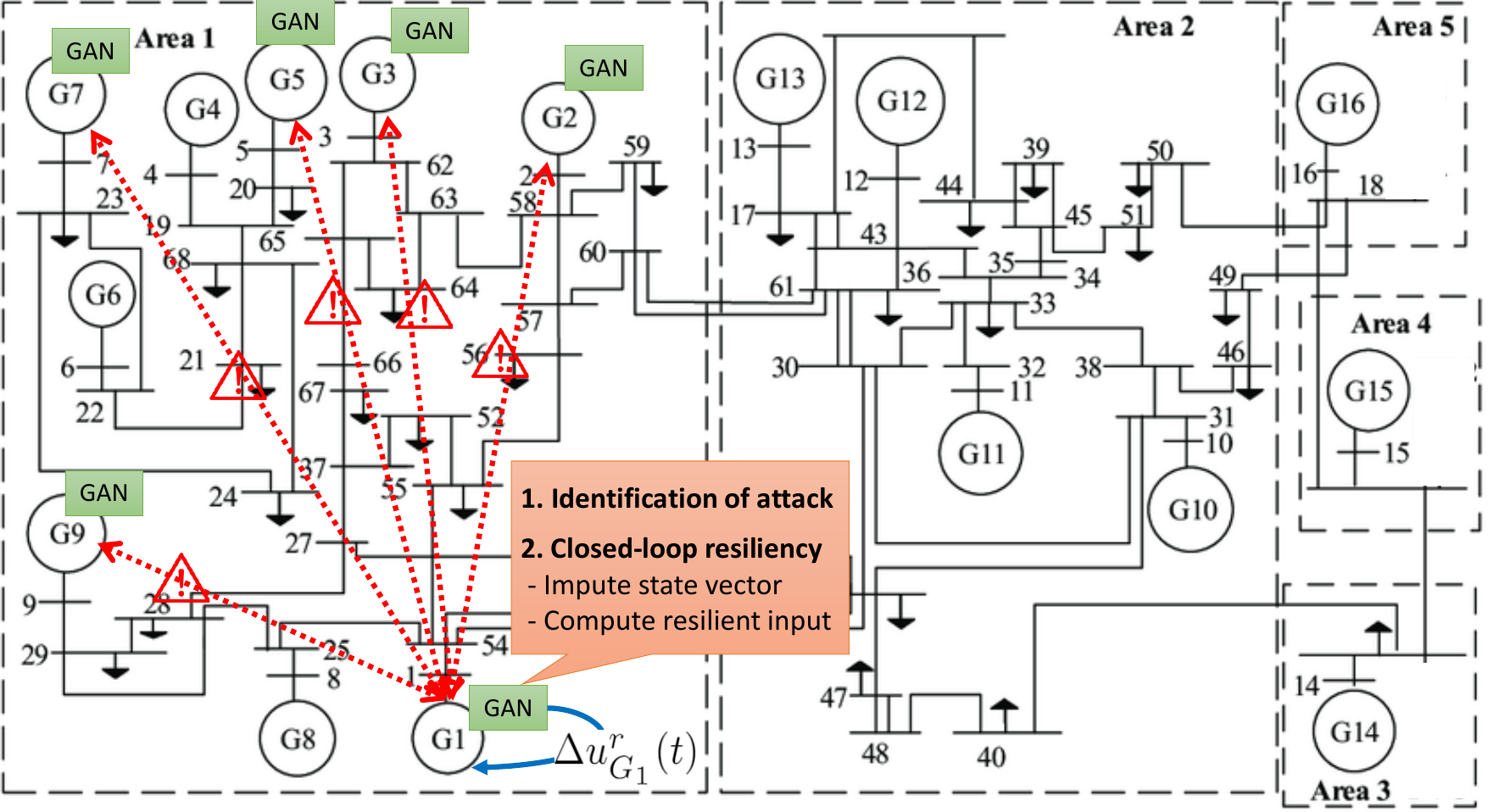}
    \caption{IEEE 68-bus, 16-machine power system model}
    \label{fig:68bus}
\end{figure}

\vspace{-0.02in}
In this section, we show the effectiveness of our proposed GAN in detecting and ensuring resiliency to FDI and DoS attacks using the IEEE 68-bus, 16-machine system as our test system as shown in Figure \ref{fig:68bus}. 
The attacked links are shown in red. PMUs are assumed to be present at all generator buses. Following literature, we assume a mean intra-area delay of $\tau = 30$ms, and an inter-area delay of $\tau = 60$ms. We assume the delay uncertainty to be $\pm 10\%$ of the mean value.
\subsection{1. Training of GAN}
The training data consists of state trajectories for $t \in [0,25]$ for operating points corresponding to load changes at buses 8, 17, 18, 23, 29, 41, 46 by a step change of $5\%$ with a maximum change of $\pm 15\%$ at any bus.  A total of 5000 distinct operating points are obtained (note that not all load variations were considered). The attacked data input for the training of the generator was created for 20 different attack scenarios $\Omega_i, i=1,2,..,20$. $\Omega_0$ was chosen as the case with no attack, i.e the generator must output the same vector as the input. For the different attack cases, the attacked entries were first set to zero (indicating a DoS attack), followed by a random bias of $20 \%$ (indicating a FDI attack). The value of $\lambda$ is chosen as 0.01, and the window size as $w=5$.
\begin{figure}[h]
    \centering
    \includegraphics[width=\linewidth,height=3cm]{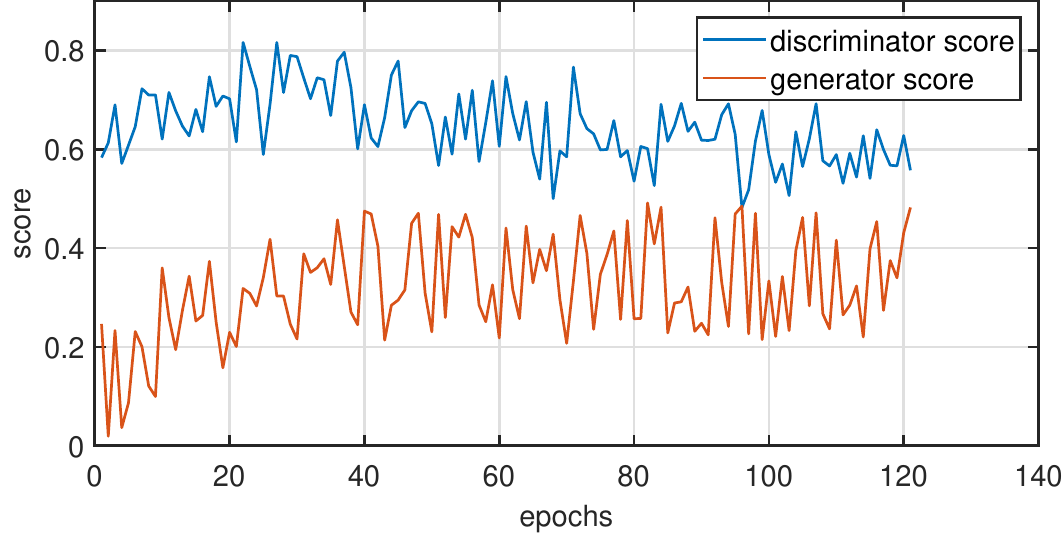}\vspace{-0.32cm}
    \caption{Generator vs discriminator score (training)}
    \label{fig:training}
\end{figure}

Fig. \ref{fig:training} shows the training results for a maximum of 120 epochs. The ADAM optimizer \cite{gan} was used during the training. The generator score is $\mathbb{E}[\mathbf{D}\big( G(\mathcal{Z}) \big)]$, and the discriminator score is $\mathbb{E}[\mathbf{D}\big( \mathcal{X}\big)]$, where $\mathcal{Z}$ is the attacked data, and $\mathcal{X}$ is the real data. We see that both scores come close to 0.5, indicating that the discriminator is no longer able to efficiently discriminate between generated and real data, which implies that the generator is well-trained \cite{gan}. 

\subsection{2. False data injection and DoS attacks}
\begin{figure}[h]
    \centering
    \includegraphics[width=0.95\linewidth,height=3cm]{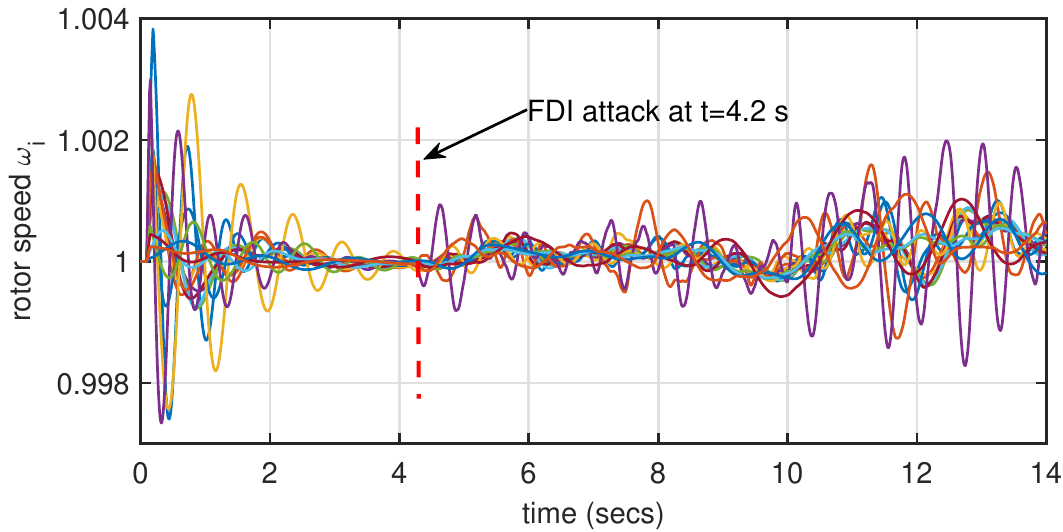}
    \caption{FDI attack destabilizing the system}
    \label{fig:fail}
\end{figure}

\vspace{-0.1cm}
Fig. \ref{fig:fail} shows the case when a random bias of $12\%$ amplitude is added to the state vector shared over the attacked communication links shown in Fig. \ref{fig:68bus}.
We see that once the attack is executed at $t=4.2$ seconds, the system slowly deviates from the damped trajectory into large oscillations towards instability. Next, we test how we can successfully identify this attack, and reduce large oscillations using the proposed GAN based prediction method.

\begin{figure}[h]
    \centering
    \includegraphics[width=0.95\linewidth,height=3cm]{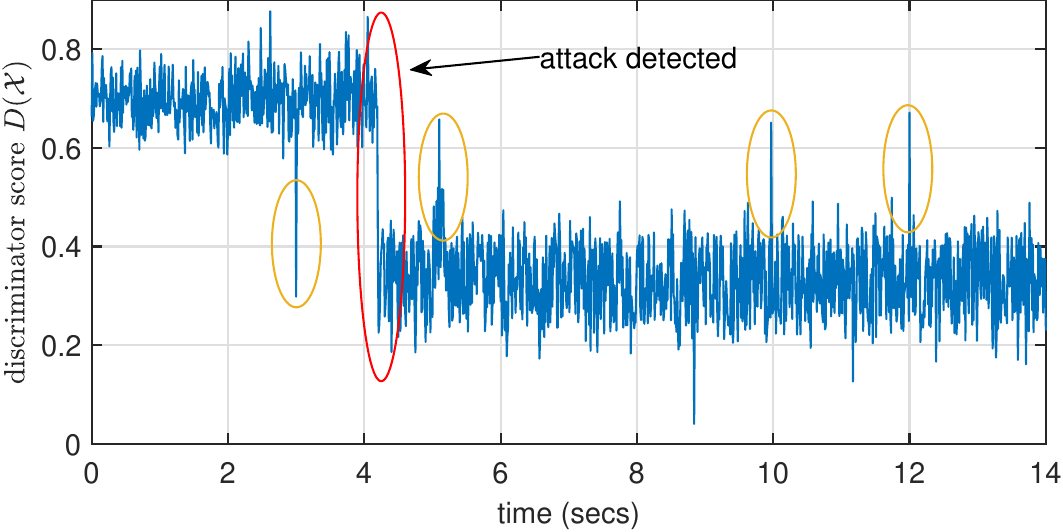}
    \caption{Attack detection using discriminator output}
    \label{fig:detection}
\end{figure}

\textbf{Identification} : Fig. \ref{fig:detection} shows the discriminator score $D(\mathcal{X})$ for the attacked data at generator 1. We can see at $t=4.2$ seconds, the score significantly drops below the threshold of $\epsilon = 0.5$, thereby indicating an attack. However, there are certain outlier values (indicated in yellow ovals) which can be attributed to training error which falsely indicate attack $(<\epsilon)$ and no-attack $(>\epsilon)$.

\begin{figure}[h]
    \centering
    \includegraphics[width=0.95\linewidth,height=3cm]{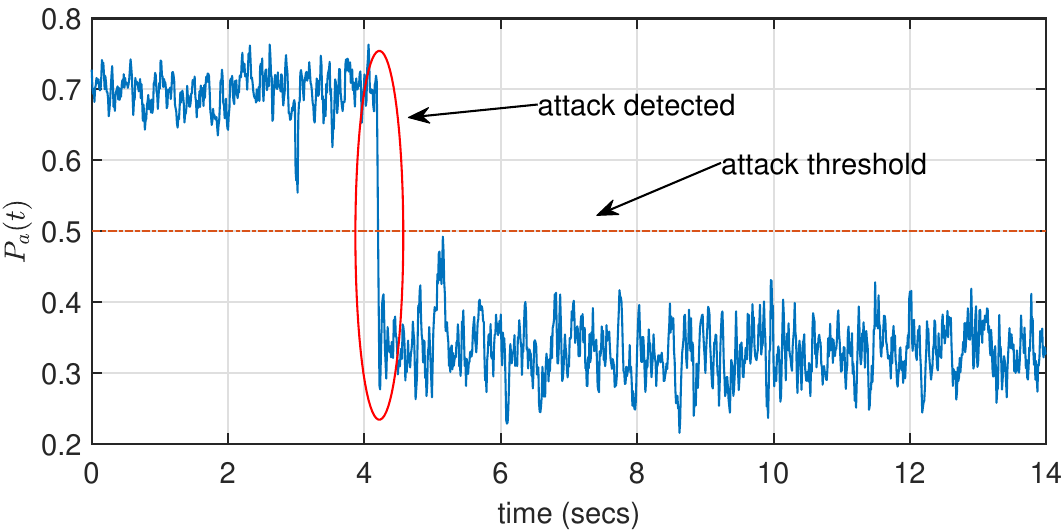}
    \caption{Attack detection using moving average}
    \label{fig:Pa}
\end{figure}

\begin{figure}[h]
    \centering
    \includegraphics[width=0.95\linewidth,height=3cm]{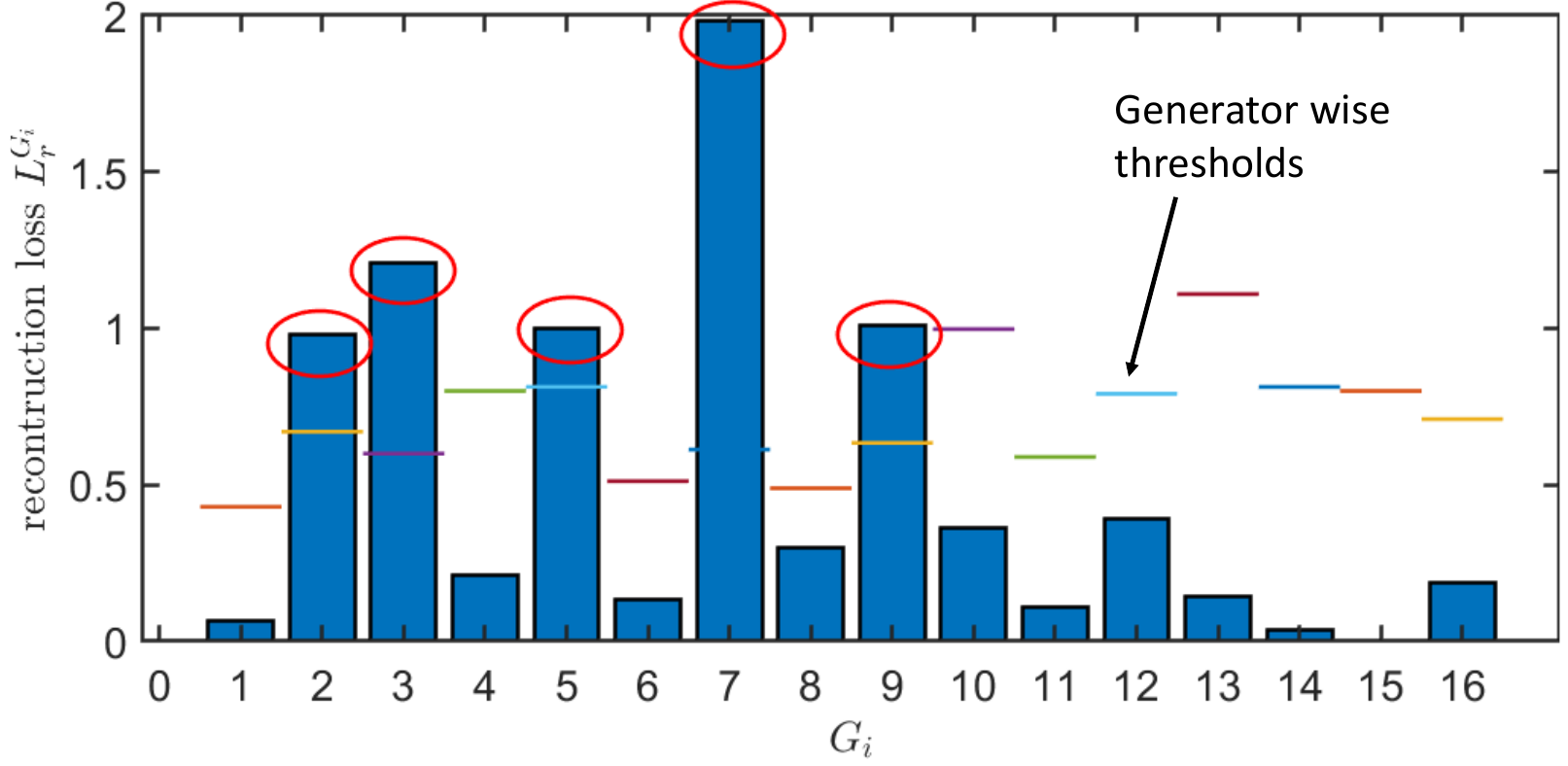}
    \caption{Attack identification using $L_r^{G_j}$}
    \label{fig:elementwisethreshold}
\end{figure}

Accordingly, in Fig. \ref{fig:Pa} we compute a moving window probability score $P_a(t)$ following (\ref{eqn:Pa}). We choose the moving window length as $d=4$. From Fig. \ref{fig:Pa} we see that the proposed moving average score eliminates the false detection without any consequential impact on the detection of the FDI attack. Note that a very small window size ($d<3$) may not improve the false detection notably, while a very large window ($d>6$) may lead to a late detection. \vspace{-0.2cm}
The results for identification of the attacked entries are presented in Fig. \ref{fig:elementwisethreshold}. As discussed before, the thresholds are estimated based on the average generator-wise reconstruction error. The thresholds have been marked by colored horizontal lines.
Whenever the reconstruction loss $L_r^{G_i}$ is more than the threshold, the corresponding entries are identified as attacked. In Fig. \ref{fig:elementwisethreshold}, we can see how the algorithm successfully identifies the attacked links from generators 2, 3, 5, 7, and 9 as their losses are higher than the marked thresholds. 
\subsubsection{Attack Mitigation :}
\begin{figure}[h]
    \centering
    \includegraphics[width=0.995\linewidth,height=3cm]{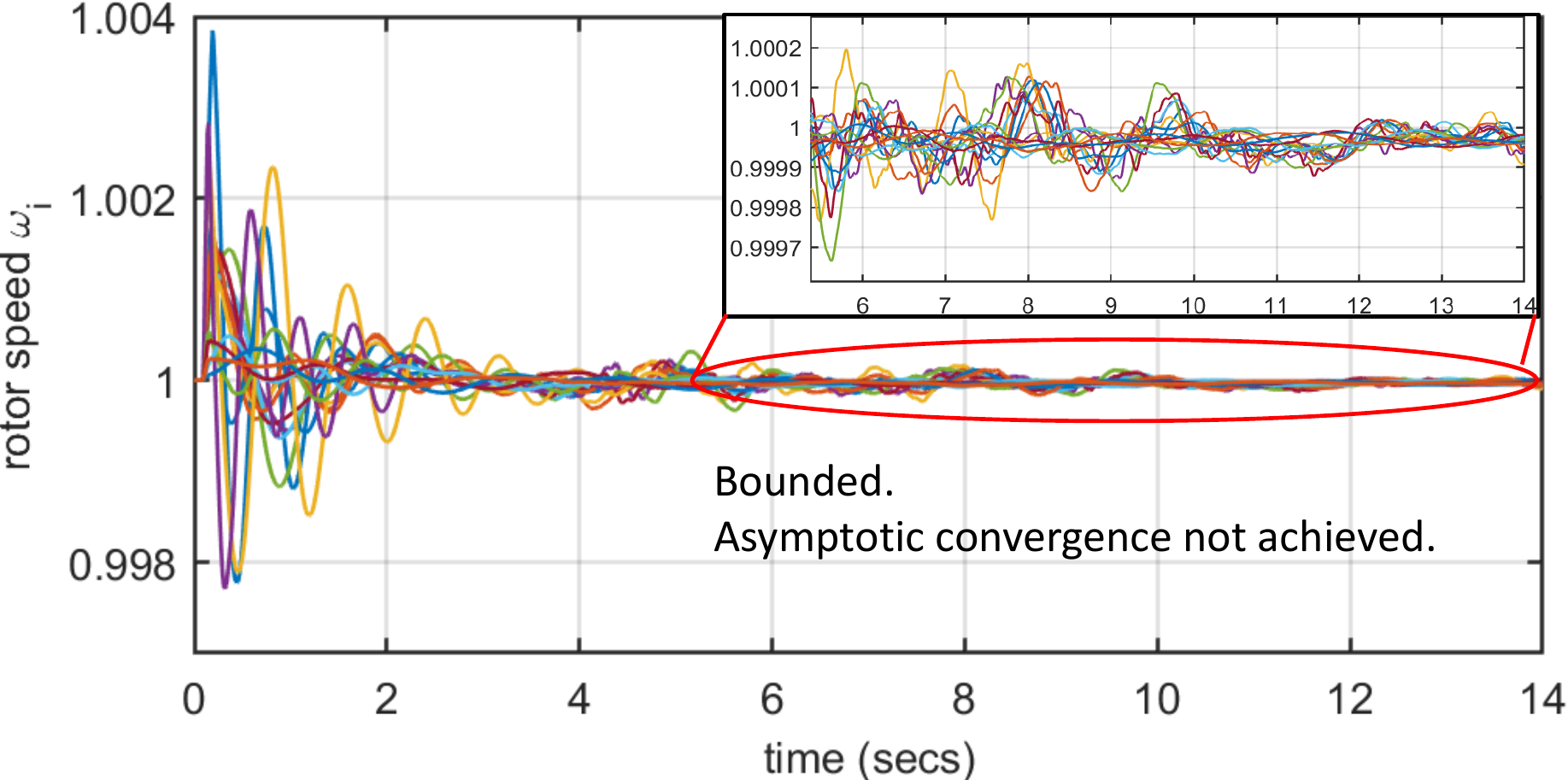}
    \caption{Resiliency against FDI attack}
    \label{fig:fdiresiliency}
\end{figure}

Fig. \ref{fig:fdiresiliency} shows the resiliency results. We can see that in Fig. \ref{fig:fail}, the large oscillations arising due to the FDI attack are successfully mitigated once the controller is redesigned following Algorithm 2.

\begin{figure}[h]
    \centering
    \includegraphics[width=0.95\linewidth,height=3cm]{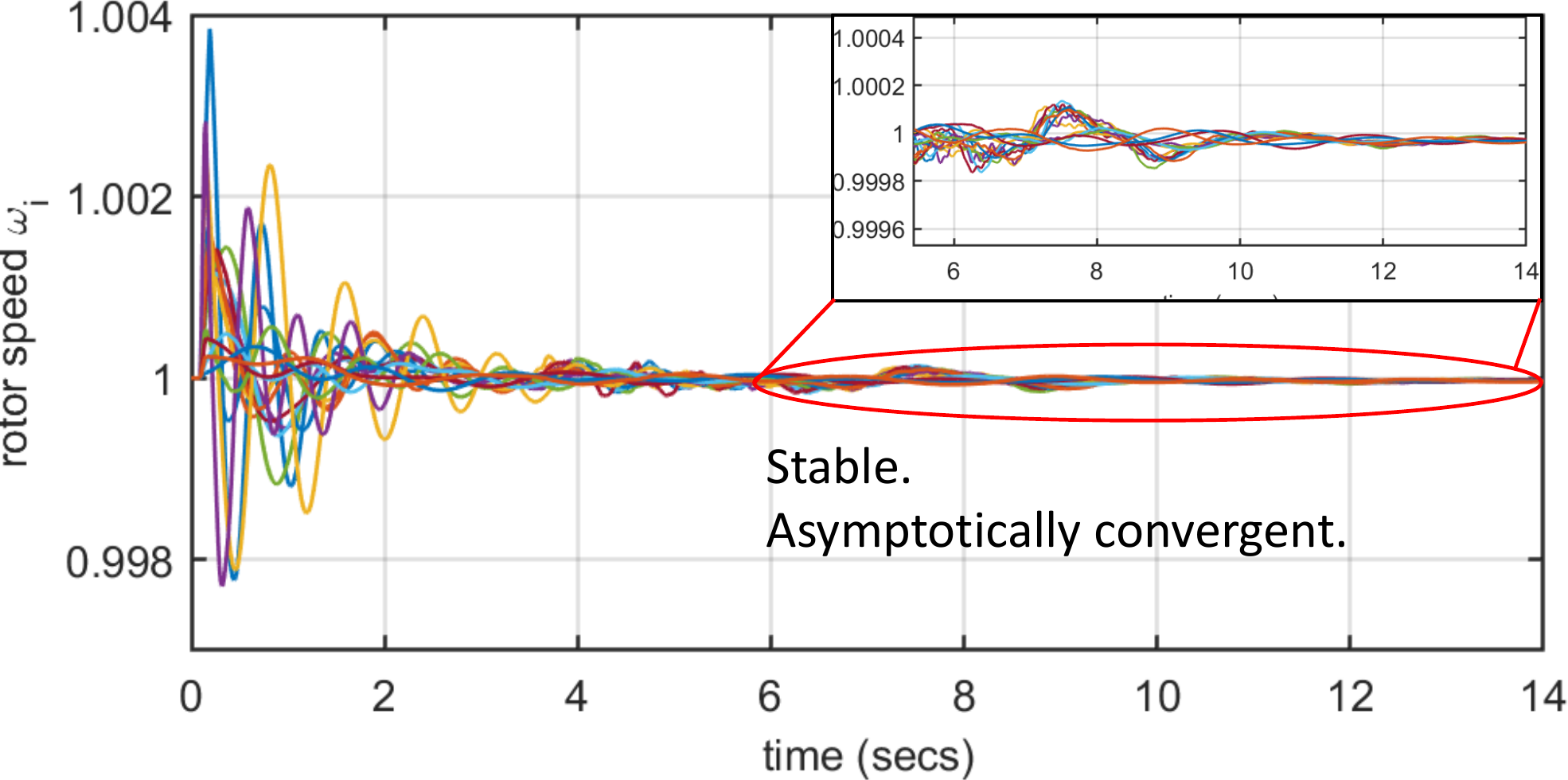}\vspace{-0.23cm}
    \caption{Resiliency against DoS attack}
    \label{fig:dosresiliency}
\end{figure}

 Fig. \ref{fig:dosresiliency}, shows the closed-loop trajectory using the proposed GAN based resiliency after a DoS attack is launched at $t=3.5$ seconds. It can be seen that states are asymptotically stable and bounded. 

\end{document}